\documentclass[prl,twocolumn,preprintnumbers,amsmath,amssymb,superscriptaddress]
{revtex4}

\usepackage[pdftex]{graphicx, color}
\usepackage{amsmath}
\usepackage{amssymb}
\usepackage{amsthm}
\usepackage{amsfonts}						%
\usepackage{graphicx}
\usepackage[english]{babel}             	
\usepackage{graphicx}                   	
\usepackage{color}

\begin{document}

\title{Observation of sub-Bragg diffraction of waves in crystals}

\author{Simon R. Huisman} \email{s.r.huisman@utwente.nl, www.photonicbandgaps.com}
\affiliation{Complex Photonic Systems (COPS), MESA+ Institute for
Nanotechnology, University of Twente, PO Box 217, 7500 AE Enschede, The Netherlands}
\author{Rajesh V. Nair} 
\altaffiliation{Present adress: Atomic and Molecular Physics Division, Bhabha Atomic Research Center, Mumbai, India}
\affiliation{Complex Photonic Systems (COPS), MESA+ Institute for
Nanotechnology, University of Twente, PO Box 217, 7500 AE Enschede, The Netherlands}
\author{Alex Hartsuiker}
\affiliation{Complex Photonic Systems (COPS), MESA+ Institute for
Nanotechnology, University of Twente, PO Box 217, 7500 AE Enschede, The Netherlands}
\affiliation{Center for Nanophotonics, FOM Institute for Atomic and
Molecular Physics (AMOLF), Science Park 113, 1098 XG Amsterdam,
The Netherlands}%
\author{L\'eon A. Woldering}
\affiliation{Complex Photonic Systems (COPS), MESA+ Institute for
Nanotechnology, University of Twente, PO Box 217, 7500 AE Enschede, The Netherlands}
\author{Allard P. Mosk}
\affiliation{Complex Photonic Systems (COPS), MESA+ Institute for
Nanotechnology, University of Twente, PO Box 217, 7500 AE Enschede, The Netherlands}
\author{Willem L. Vos}
\affiliation{Complex Photonic Systems (COPS), MESA+ Institute for
Nanotechnology, University of Twente, PO Box 217, 7500 AE Enschede, The Netherlands}

\date{\today}

\begin{abstract}

We investigate the diffraction conditions and associated formation of stopgaps for waves in crystals with different Bravais lattices. We identify a prominent stopgap in high-symmetry directions that occurs at a frequency below the ubiquitous first-order Bragg condition. This sub-Bragg diffraction condition is demonstrated by reflectance spectroscopy on two-dimensional photonic crystals with a centred rectangular lattice, revealing prominent diffraction peaks for both the sub-Bragg and first-order Bragg condition. These results have implications for wave propagation in 2 of the 5 two-dimensional Bravais lattices and 7 out of 14 three-dimensional Bravais lattices, such as centred rectangular, triangular, hexagonal and body-centred cubic.  

\end{abstract}
\maketitle


The propagation and scattering of waves such as light, phonons and electrons are strongly affected by the periodicity of the surrounding structure \cite{AshcroftMermin, Economou2010}. Frequency gaps called stopgaps, emerge for which waves cannot propagate inside crystals due to Bragg diffraction. Bragg diffraction is important for crystallography using X-ray diffraction \cite{James1958} and neutron scattering \cite{Brockhouse1995}. Diffraction determines electronic conduction of semiconductors \cite{AshcroftMermin, Economou2010} and of graphene \cite{Geim2007}, and broad gaps are fundamental for acoustic properties of phononic crystals \cite{Yang2003, Yang2004} and optical properties of photonic metamaterials \cite{Yablonovitch1987, Joannopoulos2008}.   


Bragg diffraction is described in reciprocal space by the Von Laue condition $\overrightarrow{k}_{\rm{out}}-\overrightarrow{k}_{\rm{in}}=\overrightarrow{g}$, where $\overrightarrow{k}_{\rm{out}}$, $\overrightarrow{k}_{\rm{in}}$ are the outgoing and incident wave vectors and $\overrightarrow{g}$ is a reciprocal lattice vector. As a result, a plane exists in reciprocal space for which the Von Laue condition is satisfied, called a Bragg plane. When the incident and outgoing wave vectors are located on a Bragg plane these waves are hybridized, thereby opening up a stopgap at the Bragg condition. The boundary of the Brillouin zone is formed by intersecting Bragg planes and therefore gaps open on this boundary \cite{AshcroftMermin}. When diffraction involves a single Bragg plane, we are dealing with well-known simple Bragg diffraction, which corresponds in real space with the well-known Bragg condition: $m \lambda = 2 d \cos (\theta)$. Here $m$ is an integer, $\lambda$ is the wavelength inside the crystal, $\theta$ is the angle of incidence with the normal to the lattice planes, and $d$ is the spacing between the lattice planes. A stopgap is also formed when Bragg diffraction occurs on multiple Bragg planes simultaneously, which is called multiple-Bragg diffraction \cite{Chang1984}, and is fundamental for bandgap formation \cite{vanDriel2000, Romanov2001, Economou2010}. 
Wave propagation in crystals is described along high-symmetry directions \cite{AshcroftMermin}. Multiple-Bragg diffraction has been recognized in high-symmetry directions at frequencies above the first-order simple Bragg diffraction condition: $m=1$, $\lambda=2 d$ or $\overrightarrow{k}_{\rm{out}}=-\overrightarrow{k}_{\rm{in}}=\frac{1}{2}\overrightarrow{g}$. To our knowledge, multiple-Bragg diffraction has not yet been observed at frequencies below simple Bragg diffraction \cite{voetnoot1}.

In this Letter we show that for high-symmetry directions multiple-Bragg diffraction can occur at frequencies below the first order simple Bragg condition. As a demonstration we have investigated diffraction conditions for two-dimensional (2D) photonic crystals using reflectance spectroscopy. A broad stopgap is observed below the simple Bragg condition, depending on the symmetry of the lattice. Our findings are not limited to light propagation, but apply for wave propagation in general, and therefore we anticipate similar diffraction for electrons in graphene \cite{Geim2007}, and sound in phononic crystals \cite{Yang2003, Yang2004}. 

\begin{figure}[t!]
  \includegraphics[width=8.6 cm]{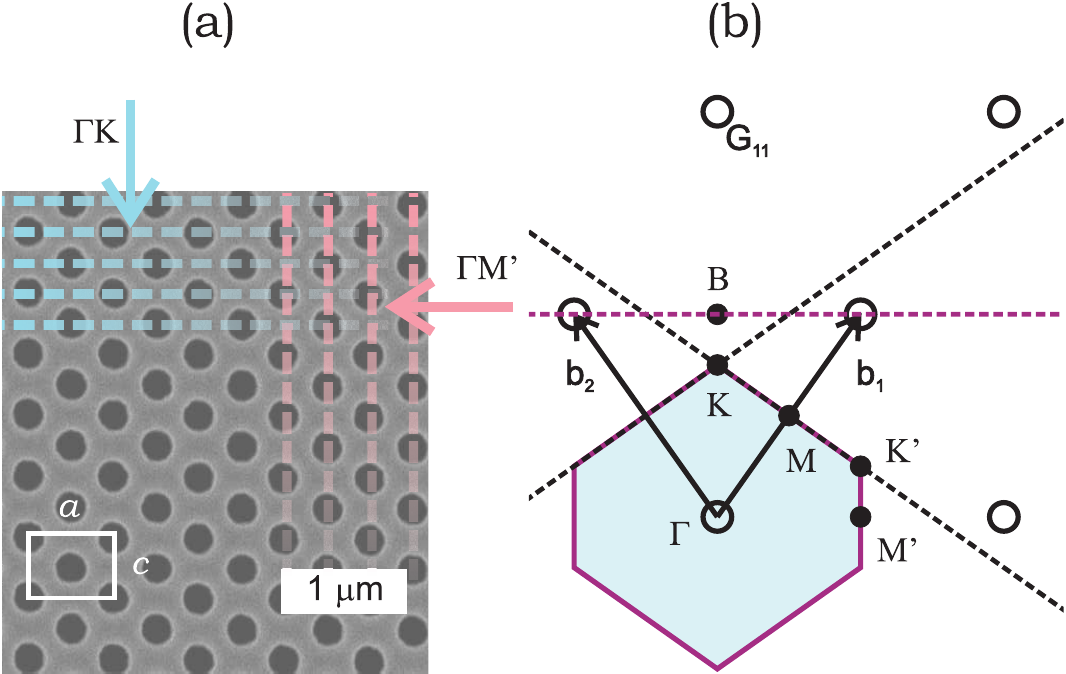}
\caption{\textit{(color online)} $(a)$ SEM image of a 2D photonic crystal with a centred rectangular lattice. The white rectangle marks an unit cell with $a=693 \pm 10$ nm, $c=488 \pm 11$ nm and $r=155 \pm 10$ nm. The blue and pink arrows mark two directions of high symmetry $\Gamma K$ and $\Gamma M^\prime$. The blue and pink dashed lines mark real space lattice planes whose lowest-frequency simple Bragg diffraction occurs along the $\Gamma K$- and $\Gamma M^\prime$-directions. $(b)$ Reciprocal space of the centred rectangular lattice (circles). The filled area is the first Brillouin zone, $b_1$ and $b_2$ are primitive vectors. $\Gamma$, $K$, $K^\prime$, $M$, and $M^{\prime}$ are points of high symmetry. The dashed lines are Bragg planes.}
\label{fig1}
\end{figure}

We have studied light propagation in 2D silicon photonic crystals \cite{Woldering2008}. Figure \ref{fig1}$(a)$ shows a scanning electron microscope (SEM) image of one of these crystals from the top view. The centred rectangular unit cell has a long side $a=693 \pm 10$ nm and a short side $c=488 \pm 11$ nm. The pores have a radius of $r=155 \pm 10$ nm and are approximately $6$ $\mu$m deep. The photonic crystals are cleaved parallel to either the $a$-side or $c$-side of the unit cell. The cleavages define two directions of high symmetry, $\Gamma M^\prime$ and $\Gamma K$, in the Brillouin zone, see Figure \ref{fig1}$(b)$. If light travels parallel with these directions, one expects simple Bragg diffraction from the lattice planes in real space (dashed lines in Figure \ref{fig1}$(a)$). A stopgap should appear that is seen in reflectivity as a diffraction peak. Because both directions are of high-symmetry, one naively expects that simple Bragg diffraction to give the lowest-frequency diffraction peak. 

We have identified the diffraction conditions of our 2D photonic crystals along the $\Gamma M^\prime$- and $\Gamma K$-directions using reflectance spectroscopy \cite{Huisman2011}.  The photonic crystals are illuminated with a supercontinuum white light source (Fianium SC-450-2). TE-polarized light is focused on the crystal using a gold-coated reflecting objective (Ealing X74) with a numerical aperture of 0.65, resulting in a spectrum angle-averaged over $0.44 \pi \pm 10 \%$ sr solid angle in air. By assuming an average refractive index (n=2.6), the angular spread inside the crystal is only 14$^o$, corresponding to $0.06 \pi \pm 10 \%$ sr solid angle. The diameter of the focused beam is estimated to be $2 w_0=1$ $\mu$m. Reflected light is collected by the same objective, and the polarization is analyzed. The spectrum is resolved using Fourier transform infrared spectroscopy (BioRad FTS-6000) with an external InAs photodiode. The spectral resolution was $15$ cm$^{-1}$, corresponding to about $10^{-3}$ relative resolution. For calibration, spectra are normalized to the reflectance spectra of a gold mirror. 

\begin{figure}[t!]
  \includegraphics[width=6.45 cm]{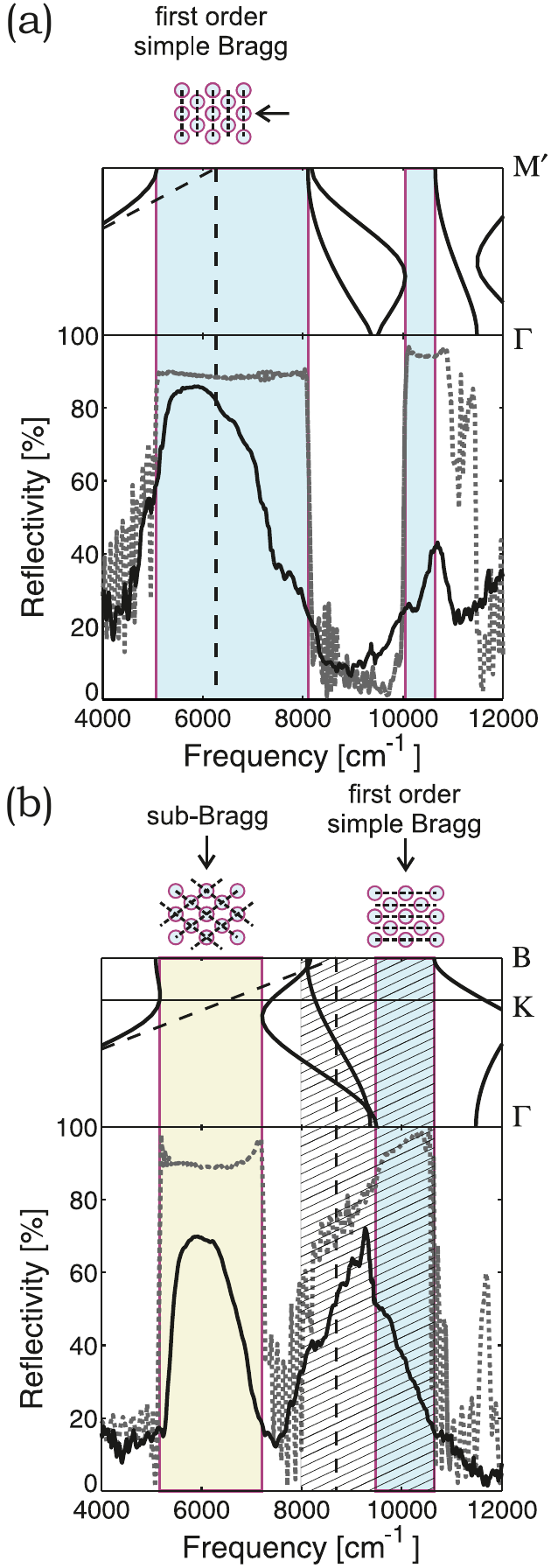}
\caption{\textit{(color online)} Measured (black, solid) and simulated (grey, dashed) reflectivity spectra, and calculated bandstructures for TE-polarized light of a 2D photonic crystal along directions of high symmetry. $(a)$ The measured and simulated lowest-frequency diffraction peaks in the $\Gamma M^\prime$-direction match a calculated stopgap that occurs at the simple Bragg-diffraction condition. $(b)$ The measured and simulated lowest-frequency diffraction peaks in the $\Gamma K$-direction match a calculated stopgap and is caused by multiple-Bragg diffraction that occurs at a lower frequency than simple Bragg diffraction. }
\label{fig2}
\end{figure}

Figure \ref{fig2}$(a)$ shows the bandstructure calculated using a plane wave expansion method \cite{mpb} and reflectivity measured along the $\Gamma M^\prime$-direction (black solid). The broad lowest-frequency measured reflectivity peak between $4700$ and $7300$ cm$^{-1}$ agrees well with the calculated stopgap. This reflectivity peak is caused by simple Bragg diffraction on the lattice planes indicated in the cartoon above, corresponding to the pink lattice planes in Figure \ref{fig1}$(a)$. One can also approximate the lowest-frequency simple Bragg diffraction condition from the dispersion with a constant effective refractive index $(n_{\rm{eff}})$, obtained from the low-frequency limit \cite{Vos1996}. This estimation is marked by the dashed vertical line and agrees well with the calculated stopgap. The two measured peaks between $9800$ and $11100$ cm$^{-1}$ agree well with a higher-frequency stopgap marked by a second blue area, caused by multiple-Bragg diffraction. The peaks appear at higher frequency than simple Bragg diffraction, as expected. The reflectivity of an incident plane wave on a finite size structure has been simulated with finite difference time domain (FDTD) simulations (grey, dashed). The agreement between the simulated and measured reflectivity is gratifying. 

In Figure \ref{fig2}$(b)$ we show the calculated bandstructure and measured reflectivity along the $\Gamma B$-direction, where $K$ is located on the edge of the Brillouin zone and $B$ is located on the Bragg-plane between $\Gamma$ and reciprocal lattice vector $G_{11}$. Two significant broad measured reflectivity peaks are visible. The lowest-frequency peak between $5400$ and $6900$ cm$^{-1}$ agrees well with a calculated stopgap marked by the yellow area. This peak is caused by multiple-Bragg diffraction on the lattice planes indicated in the cartoon above the calculated stopgap, and is part of the two-dimensional band gap for TE-polarized light. The second reflectivity peak between $8100$ and $10000$ cm$^{-1}$ agrees with a second calculated stopgap (blue area). The flat bands in the dispersion relation, causing an impedance mismatch of coupling light into the crystal \cite{Sakoda1995, Joannopoulos2008}, likely broaden the observed peak (hatched area). This is supported by FDTD simulations of the reflectivity of an incident plane wave on a finite size structure (grey, dashed). The agreement between the simulated and measured reflectivity peak is gratifying. The measured peak is probably rounded-off as a result of the high-NA microscope objective. Note that bandstructure calculations and FDTD simulations neglect the dispersion of silicon. Scattering from surface imperfections becomes more important at higher frequencies, which could explain why the measured reflectivity peak is much lower near $10000$ cm$^{-1}$. At any rate, the frequency ranges of the measured and simulated peaks agree very well.

This second stopgap is caused by simple Bragg diffraction on the lattice planes indicated in the cartoon above the calculated stopgap, corresponding to the blue lattice planes in Figure \ref{fig1}$(a)$. The frequency of the simple Bragg diffraction condition based on an $n_{\rm{eff}}$ is inaccurate because a broad stopgap is already present at lower frequencies. The observation of a prominent diffraction peak caused by multiple-Bragg diffraction at a much lower frequency than simple Bragg diffraction is important. This result shows that even for high-symmetry directions such as the $\Gamma K$-direction, there can be a diffraction condition below simple Bragg diffraction, which we address as \textit{sub-Bragg} diffraction \cite{Spikhalskii1986}  

\begin{figure}[t!]
  \includegraphics[width=8.8 cm]{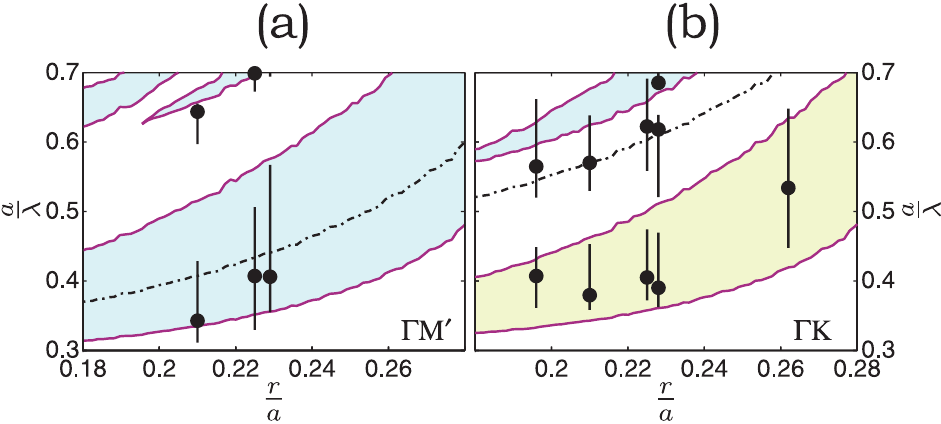}
\caption{\textit{(color online)} Determined reduced width of the diffraction peaks (bars) and frequency of the maximum reflectivity (circles) for different $\frac{r}{a}$. The filled areas are calculated stopgaps, color-coded as in Fig. \ref{fig2}. $(a)$ Reduced frequency of the diffraction peaks for the $\Gamma M^\prime$-direction. $(b)$ Reduced frequency of the diffraction peaks for the $\Gamma K$-direction.}
\label{fig3}
\end{figure}

We have performed reflectivity measurements on photonic crystals with a range of $\frac{r}{a}$. Figure \ref{fig3} shows the width of the diffraction peaks for the $\Gamma M^\prime$- $(a)$ and $\Gamma K$-directions $(b)$. The areas correspond with calculated stopgaps, such as in Figure \ref{fig2}. The dash-dotted line is the approximated frequency of lowest-frequency simple Bragg diffraction assuming a constant $n_{\rm{eff}}$. Note the very good agreement between the measured frequencies of the diffraction peaks and the calculated stopgaps. We observe for the $\Gamma K$-direction diffraction always appearing at a lower frequency than simple Bragg diffraction. This observation confirms the robustness of sub-Bragg diffraction.

The existence of sub-Bragg diffraction can be explained by considering the lattice in reciprocal space, see Figure \ref{fig1}$(b)$. For the $\Gamma K$-direction we observe in reciprocal space two points of high symmetry: $K$ and $B$. $K$ is located on the Brillouin zone boundary, at the intersection of two Bragg planes corresponding to the von Laue conditions between $\Gamma$ and $G_{10}$, $\Gamma$ and $G_{11}$. Thus at $K$ we have a multiple-Bragg diffraction condition on both Bragg planes. $B$ is located at the Bragg plane (dashed line) that satisfies the von Laue condition between $\Gamma=G_{00}$ and $G_{11}$ resulting in simple Bragg diffraction. Since $B$ is located outside the Brillouin zone, the simple Bragg condition occurs at higher frequency than the sub-Bragg condition. From this figure we describe three conditions for sub-Bragg diffraction: \textit{i}. The diffraction condition corresponds to a point on a corner edge of the Brillouin zone, giving rise to multiple-Bragg diffraction. \textit{ii}. The incident wavevector should be along a high symmetry direction, which is satisfied by considering only reciprocal lattice vectors $G_{khl}$, for which $\vert h \vert, \vert k \vert, \vert l \vert \leq 1$ or equivalent. \textit{iii}. Sub-Bragg diffraction can only occur at a lower frequency than the simple Bragg diffraction condition. 

Using these three conditions, it becomes evident that diffraction conditions for $M$ and $M^\prime$ correspond to simple Bragg diffraction for $G_{10}$ and $G_{1\overline{1}}$ respectively. $K^\prime$ satisfies criteria \textit{i} and \textit{iii}, however, it does not satisfy criterion \textit{ii}. This diffraction condition belongs to multiple-Bragg diffraction in a direction of lower symmetry, similar to the observation in Ref. \cite{vanDriel2000}. Therefore, sub-Bragg diffraction is only observed at $K$. In this case we have measured the reflectivity of photonic crystals that strongly interact with light. For our crystals, we find that for $\frac{r}{a} > 0.07$ a stopgap opens at $K$, and for $\frac{r}{a} \leq 0.07$ flat dispersion bands appear. 

\begin{figure}[t!]
  \includegraphics[width=8.8 cm]{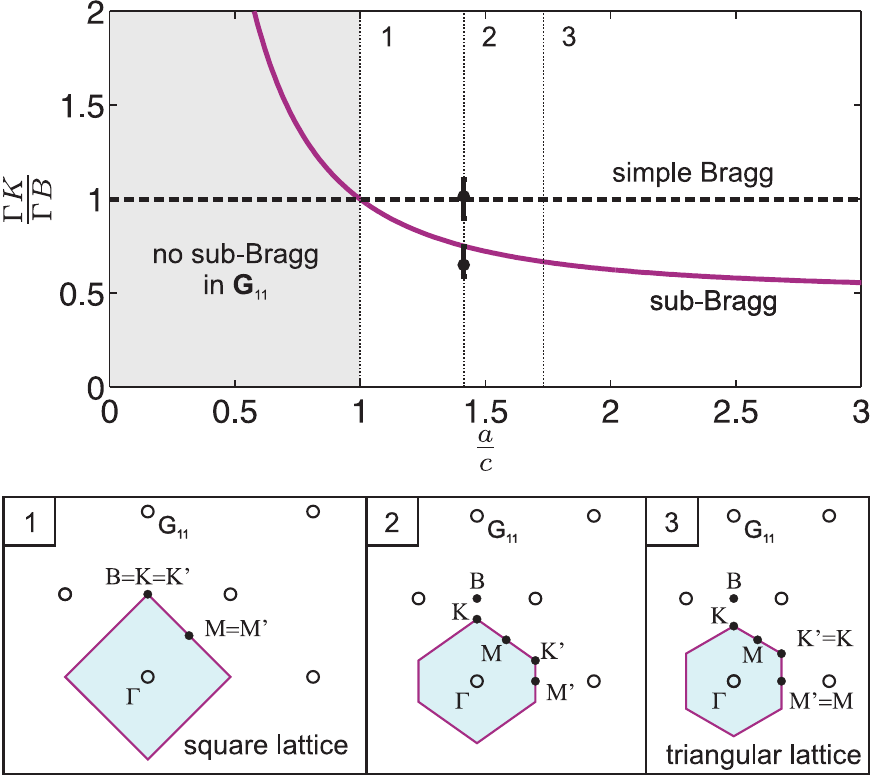}
\caption{\textit{(color online)} Normalized sub-Bragg diffraction condition (purple) and normalized simple Bragg diffraction condition (black dashed) as a function of $\frac{a}{c}$. The symbols mark the reflectivity peaks of Fig. \ref{fig2}$(b)$, assuming identical $n_{\rm{eff}}$ for both diffraction conditions. Sub-Bragg diffraction is satisfied for $\frac{a}{c}>1$. Labels $1,2,3$ refer to $\frac{r}{a}$ shown in the insets (bottom). Inset 1 shows the reciprocal lattice (circles) for $\frac{a}{c}=1$, giving a square lattice and no sub-Bragg diffraction. Inset 2 shows the reciprocal lattice (circles) for $\frac{a}{c}=\sqrt{2}$, resulting in sub-Bragg diffraction at $K$. Inset 3 shows the reciprocal lattice (circles) for $\frac{a}{c}=\sqrt{3}$, resulting in sub-Bragg diffraction at $K$ and $K^\prime$.}
\label{fig5}
\end{figure}

Up to now we have considered a centred rectangular lattice with long side $a$, short side $c$, and $\frac{a}{c}=\sqrt{2}$. However, sub-Bragg diffraction can be expected for any $\frac{a}{c}>1$ \cite{voetnoot2}. To illustrate this we have made an analytical model to explain the sub-Bragg diffraction frequency. We calculate $\vert \Gamma K \vert$ and $\vert \Gamma B \vert$ as a function of $\frac{a}{c}$, where the frequency of the sub-Bragg condition is proportional to $\frac{c_0}{n_{\rm{eff}}} \vert \Gamma K \vert$ and the frequency of the simple Bragg condition is proportional to $\frac{c_0}{n_{\rm{eff}}} \vert \Gamma B \vert$, where $c_0$ is the vacuum velocity. The results are shown in Figure \ref{fig5}. When $\frac{a}{c} \rightarrow \infty$, sub-Bragg diffraction occurs at $\frac{\vert \Gamma K \vert}{\vert \Gamma B \vert}=\frac{1}{2}$. Inset 1 shows the reciprocal lattice for $\frac{a}{c}=1$, corresponding to the square lattice. In this case $\vert \Gamma K \vert=\vert \Gamma B \vert$ and therefore sub-Bragg diffraction and simple Bragg diffraction occur at the same frequency, violating condition \textit{iii}. Inset 2 shows the reciprocal lattice for $\frac{a}{c}=\sqrt{2}$, corresponding to the experimental conditions of the structures investigated by us. Inset 3 shows the reciprocal lattice for $\frac{a}{c}=\sqrt{3}$, corresponding to the triangular lattice. All three conditions for sub-Bragg diffraction at $K$ are fulfilled. There is also a sub-Bragg diffraction condition for $K^\prime$. It may seem that condition $ii$ is violated because the $\Gamma K ^\prime$-direction corresponds to $G_{2\overline{1}}$. However, because of the rotational symmetry of the Brillouin zone, $K=K^{\prime}$ and the diffraction conditions in the $G_{2\overline{1}}$-direction are identical to the $G_{11}$-direction, and therefore condition $ii$ is satisfied. In a similar experiment performed by \cite{Rowson1999} 
a diffraction peak was observed at \textit{K}. However, these excellent experiments were compared with bandstructures between $\Gamma K$, since it was not recognized that there is also a diffraction condition at $B$. For the centred rectangular lattice, one must calculate bandstructures between $\Gamma B$ to get accurately estimate the width of the stopgaps. This is evident from the bandstructures in Figure \ref{fig2}$(b)$ by comparing the width of the stopgaps when one would consider only $\Gamma K$ instead of $\Gamma B$.

In the case of three-dimensional (3D) crystals, if a Bravais lattice has a planar cross-section that can be described by a centred rectangular lattice along a direction of high symmetry, sub-Bragg diffraction will occur. For 2D Bravais lattices sub-Bragg diffraction can occur for 2 out of 5 Bravais lattices; centred rectangular and triangular (which is a special case of centred rectangular), see Figure \ref{fig5}. There are 7 out of 14 3D Bravais lattices that have a planar cross-section that can be described by a centred rectangular lattice in a direction of high symmetry; body-centred cubic, body-centred tetragonal, base-centred orthorhombic, body-centred orthorhombic, face-centred orthorhombic, base-centred monoclinic and hexagonal. We predict that sub-Bragg diffraction can occur for these 7 Bravais lattices. 

Sub-Bragg diffraction is valid for any kind of wave-propagation in structures that fulfill the symmetry conditions. Therefore we predict that for X-ray spectroscopy on crystals a sub-Bragg diffraction peak can be observed. As multiple-Bragg diffraction is required for photonic band gap formation, hence sub-Bragg diffraction can affect band gap formation \cite{Vos2000}. Indeed, the sub-Bragg diffraction condition is part of the 2D TE-band gap in triangular lattices \cite{Joannopoulos2008}. For elastic wave diffraction a propagation gap is formed at the sub-Bragg condition and therefore also for phonons and for relativistic electrons, such as the case of graphene, which has a triangular lattice.  

We thank Willem Tjerkstra and Johanna van den Broek for expert sample fabrication and preparation, and Merel Leistikow and Georgios Ctistis for helpful discussions. This work was supported by FOM that is financially supported by NWO, and NWO-VICI, STW-NanoNed, and Smartmix-Memphis.


\begin{thebibliography}{10}
\bibitem{voetnoot2}
When $\frac{a}{c}<1$, sub-Bragg diffraction occurs in the direction perpendicular to the $\Gamma K$-direction for $\frac{a}{c}>1$. For $\frac{a}{c}<1$ it is again a centred rectangular lattice with a $90^0$ rotated unit cell compared to $\frac{a}{c}>1$.
\end{thebibliography}
\end{document}